\documentclass[twocolumn,showpacs,superscriptaddress,preprintnumbers,amsmath,amssymb,aps,pra,10pt]{revtex4-1}

\usepackage{graphicx}
\usepackage{dcolumn}
\usepackage{bm}
\usepackage{color}
\usepackage{inputenc} 
\usepackage{csquotes}

\begin{document}
\title{Doppler effect as a tool for ultrashort electric field reconstruction.}

\author{P. B\'ejot}
\affiliation{Laboratoire Interdisciplinaire Carnot de Bourgogne, UMR CNRS 6303 - Universit\'e Bourgogne Franche-Comt\'e, 21078 Dijon Cedex, France}
\author{E. Szmygel}
\affiliation{Laboratoire Interdisciplinaire Carnot de Bourgogne, UMR CNRS 6303 - Universit\'e Bourgogne Franche-Comt\'e, 21078 Dijon Cedex, France}
\affiliation{Femto Easy, Batiment Gienah, Cit\'e de la Photonique, 11 avenue de Canteranne, 33600 Pessac France.}
\author{A. Dubrouil}
\affiliation{Femto Easy, Batiment Gienah, Cit\'e de la Photonique, 11 avenue de Canteranne, 33600 Pessac France.}
\author{F. Billard}
\affiliation{Laboratoire Interdisciplinaire Carnot de Bourgogne, UMR CNRS 6303 - Universit\'e Bourgogne Franche-Comt\'e, 21078 Dijon Cedex, France}
\author{B. Lavorel}
\affiliation{Laboratoire Interdisciplinaire Carnot de Bourgogne, UMR CNRS 6303 - Universit\'e Bourgogne Franche-Comt\'e, 21078 Dijon Cedex, France}
\author{O. Faucher}
\affiliation{Laboratoire Interdisciplinaire Carnot de Bourgogne, UMR CNRS 6303 - Universit\'e Bourgogne Franche-Comt\'e, 21078 Dijon Cedex, France}
\author{E. Hertz}
\email{edouard.hertz@u-bourgogne.fr}
\affiliation{Laboratoire Interdisciplinaire Carnot de Bourgogne, UMR CNRS 6303 - Universit\'e Bourgogne Franche-Comt\'e, 21078 Dijon Cedex, France}

\begin{abstract}
Technological advances in femtosecond laser sources call for the development of increasingly refined characterization tools implying to enrich the existing panel of operable nonlinear interactions. The present paper describes a novel characterization method based on spectral-shearing interferometry which exploits the non-standard \textit{rotational Doppler effect} for producing the frequency shear. This original approach, called DEER \enquote{Doppler Effect for Electric field Reconstruction}, provides a spectral-shearing in the absence of frequency conversion and features multiple benefits as for instance the ability of operation in the ultraviolet spectral range or the characterization of ultra-broadband laser pulses. 
\end{abstract}



\newlength{\textlarg}
\newcommand{\strike}[1]{%
  \settowidth{\textlarg}{#1}
  #1\hspace{-\textlarg}\rule[0.5ex]{\textlarg}{0.5pt}}

\maketitle

\section{Introduction}
The Doppler effect~\cite{Doppler1842} is mostly known in its linear form, also called \enquote{Translational Doppler Effect}, by which the relative velocity between a wave source and an observer causes a frequency shift readily understood from the conservation of linear momentum and energy during the light-matter interaction. The translational Doppler effect is widely used for various applications such as laser remote sensing, laser velocimetry, health care, satellite communication, and robotics \cite{DurstHoweRichter1982,Lambourne1997,Liu31Oct.-3Nov.1999,AgarwalGauravNiralaEtAl2018}. In astronomy, the discovery of exoplanets by Doppler spectroscopy was awarded by the Nobel prize in physics in 2019 \cite{MayorQueloz1995}. Another less well-known manifestation of the Doppler effect is the \enquote{rotational Doppler effect} which has been the subject of a growing interest in the last decade. It is the rotational counterpart of the translational Doppler effect where the frequency shift arises from the emission or absorption of a circularly polarized light by a rotating body. 
The shift is explained in this case from the conservation of angular momentum and energy. The rotational Doppler  effect manifests itself through various physical processes such as rotational scattering~\cite{Garetz1981}, photoelectron spectroscopy for probing molecular orbitals anisotropy~\cite{MiaoTravnikovaGelmukhanovEtAl2015} and has been used in satellite-based GPS navigation systems~\cite{Ashby2003} or to observe coherently spinning molecules and measure their angular speed~\cite{KorechSteinitzGordonEtAl2013,KarrasNdongHertzEtAl2015,ProstHertzBillardEtAl2018}. We demonstrate here its relevance in a radically different field of applications, namely the ultrashort pulse characterization. The generation of ultrashort pulses has known during the last decades major technological breakthroughs~\cite{RulliereBook} with the emergence of new coherent sources in the mid-infrared and ultraviolet (UV), the generation of super-continua~\cite{AlfanoShapiro1970} and near-single cycle pulses like Terahertz~\cite{HiroriDoiBlanchardEtAl2011} or XUV attosecond pulses~\cite{CalegariSansoneStagiraEtAl2016}. Theses developments have been a driving element for spectacular advances. The utilization of such pulses requires a comprehensive characterization of the electric field that led to a growing interest in reliable and versatile diagnostic tools~\cite{RulliereBook,MonmayrantWeberChatel2010}. Nowadays, the two most widespread self-referenced complete characterization methods are the frequency-resolved optical gating~\cite{TrebinoW.DeLongFittinghoffEtAl1997} (FROG) and the spectral phase interferometry for direct electric field reconstruction~\cite{IaconisWalmsley1998,AndersonMonmayrantGorzaEtAl2008} (SPIDER). The approach described in the present paper can be seen as a variant of the SPIDER. The SPIDER relies on the spectral interferometry between the pulse to be characterized and its replica that is  both spectrally and temporally shifted. The interferogram yields the gradient of the spectral phase from which the spectral phase can be derived. The frequency shift (or shear) $\Omega$ is produced by nonlinear effect. In the original version of the SPIDER, the input pulse is split in two parts, one being used to produce a highly chirped pulse and the other one to generate a sequence of two time-delayed pulses. The two replica are then sum frequency mixed in a nonlinear crystal with the chirped pulse. Since each pulse overlaps with a different quasi-monochromatic temporal slice of the chirped pulse, the generated pulses have slightly different central frequency (close to the second harmonic). The frequency shear $\Omega$ can be adjusted through the chirp and the delay between the two replica. Spectral-shearing interferometry techniques as SPIDER feature several advantages. Among them, they provide a self-referenced measurement under single-shot operation and the inversion procedure to recover the spectral phase is straightforward, non-iterative, and unambiguous since it relies on Fourier transforms and spectral filtering. Furthermore, the data processing of the SPIDER signal is simpler and faster than the one implemented with a bi-dimensional \enquote{spectrogram} obtained from a FROG or a D-scan \cite{LoriotGitzingerForget2013} techniques. The SPIDER is also rather insensitive to the spectral dependence of the nonlinear optical process and to the noise~\cite{AndersonMonmayrantGorzaEtAl2008} since the relevant information is encoded into the fringes spacing of the interferogram and not in the spectral amplitude. Since the first demonstration of SPIDER~\cite{IaconisWalmsley1998}, considerable efforts have been made to extend the scope of measurements~\cite{AndersonMonmayrantGorzaEtAl2008} giving birth to several variants of the original design enabling for instance the single-shot measurement of weak pulses~\cite{HirasawaNakagawaYamamotoEtAl2002}, the evaluation of spatio-temporal distortions~\cite{KosikRadunskyWalmsleyEtAl2005}, or measurements free from phase distortions induced by the beamsplitter~\cite{BaumLochbrunnerRiedle2004}. In all theses variants, the shear is produced by sum-frequency generation where the absorption and phase matching bandwidth of the crystal may limit the functionality of standard devices when considering ultra-broadband or ultraviolet (UV) laser pulses (besides the crystal limitation, if the wavelength is below $\approx$ 380~nm, the harmonic signal lying in the vacuum UV will be also absorbed by oxygen). A scheme called DC-SPIDER~\cite{LonderoAndersonRadzewiczEtAl2003}, relying on the difference frequency generation between the pulse to characterize and a chirped pulse of smaller frequency, has been proposed for the characterization of UV pulses. Nevertheless, the method is not \textit{strictly speaking} self-consistent and the need of the \enquote{extra} chirped pulse of smaller frequency limits the use of this approach to custom developments. Another method called SD-SPIDER has recently been reported~\cite{BirkholzSteinmeyerKokeEtAl2015}. Unlike most variants of the SPIDER, the method is based on a third-order nonlinear optical process, namely the self-diffraction (SD) effect rather than sum-frequency generation. A spectral-shearing interferometry method based on SD is in principle well suited for the characterization of UV or broadband laser pulses since it operates at the fundamental wavelength and does not acquire frequency conversion. Unfortunately, the SD process is not phase-matched and the device does not measure the phase of the pulse but that of its second harmonic requiring an elaborate retrieval procedure. Besides these drawbacks, the use of third-order nonlinear effect through spectral interferometry appears very desirable for the characterization of UV or ultra-broadband laser pulses. To this end, new concepts have to be considered and we present here an original approach exploiting the rotational Doppler frequency shift~\cite{Garetz1981,KorechSteinitzGordonEtAl2013,KarrasNdongHertzEtAl2015} that lifts the flaws inherent to the SD phenomenon. As detailed below, the overall mechanism also relies on a third-order nonlinear optical effect. Besides a perfect phase-matching and direct phase retrieval, the method features several potential advantages such as an extended functionality toward UV spectral range, spectral domain of importance since many organic systems have their absorption bands in this range. Additionally, the device only needs one spectrometer instead of two as in a standard SPIDER design, which might even be unnecessary by implementing a SEA-SPIDER configuration~\cite{KosikRadunskyWalmsleyEtAl2005}.  
Furthermore, interferometric techniques have been used for the characterization of some of the shortest pulses generated so far and methods based on a virtually unlimited bandwidth $\chi^{(3)}$ process are particularly pursued when spectrum exceeds one optical octave ~\cite{BirkholzSteinmeyerKokeEtAl2015}. 

The proof-of-principle of the present method, called DEER (Doppler Effect for Electric field Reconstruction), is conducted with a laser delivering Fourier Transform Limited (FTL) pulses of 100~fs duration at a central wavelength of 800~nm. The performance of the approach is assessed by characterizing various spectral phases applied by a standard pulse shaper device.  
 
\section{Principle}
As already mentioned, the SPIDER consists of measuring the interferogram associated to a sequence of two pulses delayed by $\tau$ and spectrally shifted by $\Omega$. Writing $\omega$ the angular frequency and $\left|E(\omega)\right|e^{i\varphi(\omega)}$ the complex electric field to be characterized (expressed in the frequency-domain), the interferogram writes as:
\begin{eqnarray}
S(\omega)&=&\left|E(\omega)\right|^2+\left|E(\omega+\Omega)\right|^2 \nonumber\\
&+&2\left|E(\omega)\right| \left|E(\omega+\Omega)\right| \textrm{cos}(\omega\tau+\Phi_S(\omega)),
\end{eqnarray}
where $\Phi_S(\omega)=\varphi(\omega+\Omega)-\varphi(\omega)$ contains the phase $\varphi(\omega)$ to be
recovered. The relevant information is encoded in the shape of the fringes. With no chirp [$\Phi_S(\omega)=0$] the argument of the cosine reduces to $\omega\tau$ and the fringes present an equal frequency spacing inversely proportional to the delay $\tau$. Any change of this interferogram (spreading or shrinking) reflects the presence of a chirp. The cosine argument can be extracted through spectral filtering and the knowledge of $\tau$ allows the evaluation of $\Phi_S(\omega)$ that is proportional to the spectral phase derivative. The spectral phase $\varphi(\omega)$ of the pulse can thus be retrieved (up to a constant) by integrating $\Phi_S(\omega)$: 
\begin{equation}
\varphi(\omega)\approx \frac{1}{\Omega}\int_{0}^{\omega} \Phi_S(\omega') d\omega'.
\label{Eq_Phase_retrival}
\end{equation}
The central key of the procedure is the production of the frequency shear $\Omega$ and we propose in this issue to exploit the rotational Doppler effect. The rotational Doppler effect can be seen as a variant of the well-known translational Doppler effect which appears for instance when a wave is reflected on a moving object. The reflected (or counter-propagating) wave is frequency shifted by $\Delta\omega=-2\vec{k}.\vec{v}$ with $\vec{k}$ the wavevector of the incident field and $\vec{v}$ the velocity of the object. The rotational Doppler effect for its part takes place for instance when a circularly polarized (CP) light interacts with an anisotropic object in rotation (in the plane of the field polarization). The counter-rotating wave is then shifted by $\Delta \omega=\pm 2~\Omega_{\textrm{rot}}$, where $\Omega_{\textrm{rot}}$ stands for the angular frequency of the anisotropic object and where the sign of the shift depends on the helicity of the CP wave with respect to the sense of rotation of the object. The frequency is downshifted (upshifted) for an incident CP field rotating in the same (opposite) sense as the object. This effect can be readily understood by considering the example of the half-wave plate~\cite{A.GaretzArnold1979}. A half-wave plate rotates the polarization direction of a linearly-polarized light by twice the angle between the incident polarization direction and the fast axis of the half-wave plate, namely from $\theta_{\textrm{in}}$ to $\theta_{\textrm{out}}=2\theta_{\textrm{wav}}-\theta_{\textrm{in}}$, with $\theta_{\textrm{wav}}$ the angle of the fast axis. The half-wave plate flips therefore the helicity of a CP wave since at any time it produces the symmetric of the electric field with respect to its fast axis ($\theta_{\textrm{in}}=-\omega t$ becomes $\theta_{\textrm{out}}=2\theta_{\textrm{wav}}+\omega t$). A rotation of the fast axis of the wave plate by $\theta_{\textrm{wav}}=\Omega_{\textrm{rot}}t$ induces therefore a variation of the resulting field vector angular speed, namely of the angular frequency by $2\Omega_{\textrm{rot}}$. It can be shown that a wave plate, other than a half-wave plate, produces a spectrally shifted counter-rotating wave together with an unshifted co-rotating wave. The same occurs with the translational Doppler effect where a wave interacting with a semi-transparent object gives rise to a spectrally shifted reflected wave with an unshifted transmitted one. Rotational doppler shifts have been observed with anisotropic optical elements in mechanical rotation~\cite{A.GaretzArnold1979} or with rotating molecules~\cite{KorechSteinitzGordonEtAl2013,KarrasNdongHertzEtAl2015,ProstHertzBillardEtAl2018}. Recent investigations~\cite{LiZentgrafZhang2016,FaucherProstHertzEtAl2016} have shown that $n$th-order harmonic generation with spinning objects provides shifts of $(n \pm 1)\Omega_{\textrm{rot}}$. Here we show that the rotational Doppler effect can also be applied for spectral-shearing interferometry. For a proper pulse characterization, the frequency shear should be a reasonable portion of the pulse's spectral bandwidth, in practice about a tenth of the spectral bandwidth is well-suited. In our case, this corresponds to a shear $\delta\lambda\approx$1~nm at 800~nm which corresponds to a frequency shear $\Omega\approx$3 10$^{12}$ rad/s or an angular speed of the anisotropy $\Omega_{\textrm{rot}}=\Omega/2=$1.5  10$^{12}$ rad/s.  Needless to say that such a rotating anisotropy cannot be reached by a mechanical rotation. Instead, our strategy relies on a nonlinear optical techniques and more precisely on a third-order instantaneous electronic Kerr effect. A linearly polarized field interacting with a piece of glass induces a birefringence proportional to the square of the applied electric field. The material acts therefore as a wave plate with neutral axes along and perpendicular to the field polarization. The requisite rotation of anisotropy, for our purpose of rotational Doppler shift, is produced by means of a pulse of twisted linear polarization (TLP), namely a pulse with a linear polarization that continuously rotates at a speed in the THz range. Such a polarization shaping has been successfully applied to induce unidirectional rotational motion of linear molecules~\cite{KarrasNdongHertzEtAl2015,ProstHertzBillardEtAl2018}. 
In the present case, the frequency shear $\Omega$ induced by electronic Kerr effect can be retrieved by calculating the nonlinear polarization revealing a counter-rotating CP field that writes as:
\begin{equation}
E(t)\propto \chi^{(3)}_{xxxx} \left|\varepsilon_{\textrm{TLP}}(t)^2\right| \varepsilon(t) e^{i (\omega_0  \pm 2 \Omega_{\textrm{rot}})t},
\end{equation}
 with $\chi^{(3)}_{xxxx}$ the third-order susceptibility, $\omega_0$ the central angular frequency, $\varepsilon_{\textrm{TLP}}(t)$ and $\varepsilon(t)$ the complex field envelope of the TLP and the input pulse, respectively.

\section{Experimental set-up}

\begin{figure*}[htpb]
\begin{center}
\includegraphics[width=17 cm]{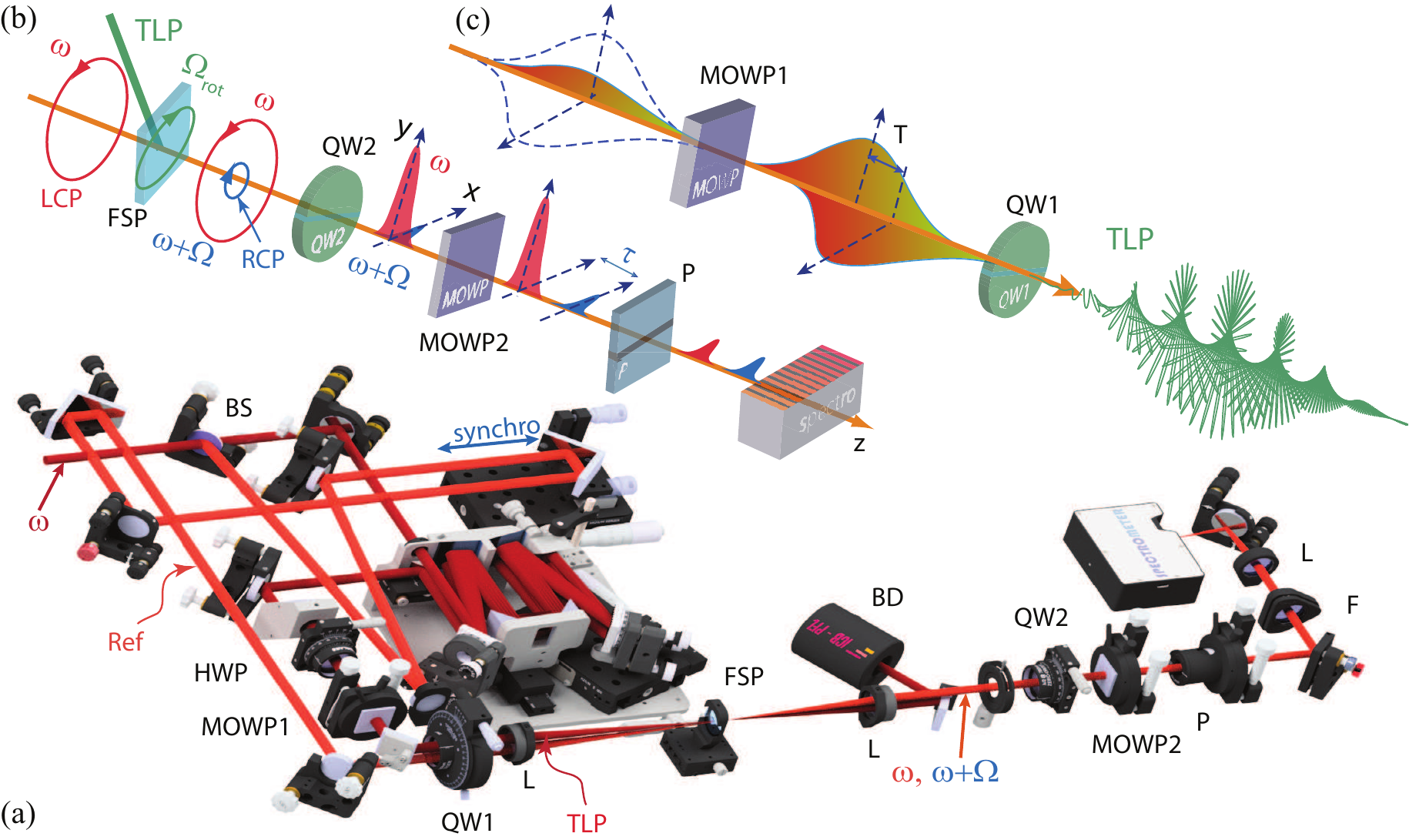}
 \caption{(a) Experimental set-up. TLP: pulse of twisted linear polarisation, BS: beam splitter, QW: quarter wave plate, MOWP: multiple order wave plate, FSP: fused Silica plate, BD: beam dump, P: polarizer, L: lens, F: neutral filter. (b) Schematic description of the spectral shearing interferometry by rotational Doppler effect (see text). (c) Scheme for the production of the TLP pulse.  }
\label{FigDetection} 
\end{center}
\end{figure*}

The DEER set-up is depicted in Fig.~\ref{FigDetection}(a). 
The input pulse (horizontally polarized) is split into two parts. One is used to produce the TLP pulse and the other one contains the \enquote{reference pulse} to analyze. Different strategies can be implemented for the production of ultrashort TLP pulses~\cite{FaucherHertzLavorelEtAl2018}. We have selected a method that relies on the production of two counter-rotating CP and time-delayed chirped pulses. Basically, a field of spinning linear polarization can be seen as the superposition of two CP fields with slightly different frequencies and opposite handedness. Our approach, depicted in Fig.~\ref{FigDetection}(c), is based on this principle. The TLP pulse is created by generating two orthogonally polarized chirped pulses that are time-delayed by $T$,  $T$ being small enough to insure a significant overlap between the two pulses. To that end, the input pulse is first temporally chirped using a double pass 4$f$-line reflective pulse stretcher. The pulse then propagates through a half wave plate (HWP) to set its polarization at 45$^{\circ}$ and next through a multiple order wave plate (MOWP1 with neutral axes along $\vec{x}$ and $\vec{y}$) providing the sequence of two time-delayed and cross polarized fields. A zero-order quarter wave plate (QW1) transforms the two perpendicular components into two CP fields of opposite helicity. Each pulse presents a frequency drift with the time so that the delay $T$ between them is equivalent to have an effective constant frequency shift. The configuration is therefore analog to the sum of two spectrally-shifted CP fields of opposite handedness resulting in the desired TLP pulse. It has been shown~\cite{FaucherHertzLavorelEtAl2018} that the field vector rotates with a temporal period $T_{\textrm{rot}}=2\pi/(aT)$ where  $a$ depends on the quadratic spectral phase $\Phi''$ applied with the stretcher:
\begin{equation}
a=\frac{\Phi''}{\frac{\Delta t^4}{(2\sqrt{2}\textrm{ln}2)^2}+2\Phi''^2},
\end{equation}
with $\Delta t$ the FTL pulse duration (full width at half maximum) considering Gaussian pulse. The main advantage of this strategy, in comparison with the one implemented so far for producing spinning molecules~\cite{KarrasNdongHertzEtAl2015}, relies on the production of longer TLP pulses with potentially many turns of the electric field (see Fig.~\ref{FigDetection}(c)). The angular Doppler effect will be therefore less sensitive to the temporal synchronization between the TLP pulse and the pulse to be spectrally shifted. In our case, the pulse of typical energy 5$\mu$J is stretched to about $\Delta t_{\textrm{ch}}=$4.4~ps ($\Phi''=$~160000~fs$^2$) and the delay is fixed to $T=500$~fs. The rotational period of the spinning polarization is $T_{\textrm{rot}}=$4~ps providing a calculated shear $\delta\lambda\approx 1.1$~nm confirmed experimentally (Fig.~\ref{FigShear}). Other configurations with shorter TLP pulses have been tested but the pulse with $\Delta t_{\textrm{ch}}=$4.4~ps is less sensitive to the input chirp and allow the characterization of spectral phases of larger amplitudes. We emphasize that the quadratic chirp, produced here by means of a pulse stretcher, could be generated using a thick dispersive medium when the pulse to characterize presents a broadband spectrum leading to a major simplification of the device. Pulse shaper devices could also offer a greater flexibility and compactness of the optical arrangement~\cite{FaucherHertzLavorelEtAl2018}. 

The  reference pulse (in the arm reflected by the beam splitter) is time-delayed with a delay line so as to be temporally synchronized with the TLP pulse in the interaction region. The reference pulse is then CP by passing through the same quarter wave plate (QW1) as the TLP pulse (Fig.~\ref{FigDetection}(a)). Reference and TLP pulses are focused by the same lens (focal length $f=$15~cm) in a fused silica plate with a small angle ($\approx4^{\circ}$). The TLP pulse induces a rotating birefringence within the plate, and the field to be characterized contains after the interaction a spectrally-shifted CP field of opposite handedness $E(\omega+\Omega)$ [or $E(\omega-\Omega)$] together with a stronger unshifted CP field of same helicity $E(\omega)$ [Fig.~\ref{FigDetection}(b)]. The spectral interferometry is conducted between these two components. The TLP beam is blocked while the reference pulse passes through a quarter wave plate (QW2). The two CP components are converted into linearly polarized fields of crossed polarization. They are further time-delayed by $\tau=3$~ps by means of a multiple order wave plate (MOWP2). A polarizer (P) is then orientated to balance their amplitudes so as to maximize the visibility of the spectral fringes measured by a spectrometer. The condition of two time-delayed sheared pulses (but otherwise identical) for spectral shearing interferometry  is therefore fulfilled in a simple and compact manner. The set of quarter wave plate (QW2) and polarizer (P) acts as a circular analyzer. We point out that any small deviations of phase retardation in the quarter wave plates QW1 and QW2 (used for the circular polarizer and analyzer of the reference pulse) has no effect as long as the two wave plates produce the same retardation. Furthermore, a particular attention has been paid to the depolarization arising from metallic reflexion. The optical set-up has been thus designed so as to only have reflections with $s$ or $p$ polarization in order to maintain the polarization.
In order to assess the quality of our device, the input pulse is spectrally phase-shaped via a standard pulse shaper device using a programmable 1D dual mask LC-SLM array (SLM-320d from Jenoptik). The applied spectral phases are compared to the measured ones. A description of the pulse shaper arrangement can be found in Ref.~\cite{HertzBillardKarrasEtAl2016}.
\section{Results}
\begin{figure}[htpb]
\begin{center}
\includegraphics[width=\linewidth]{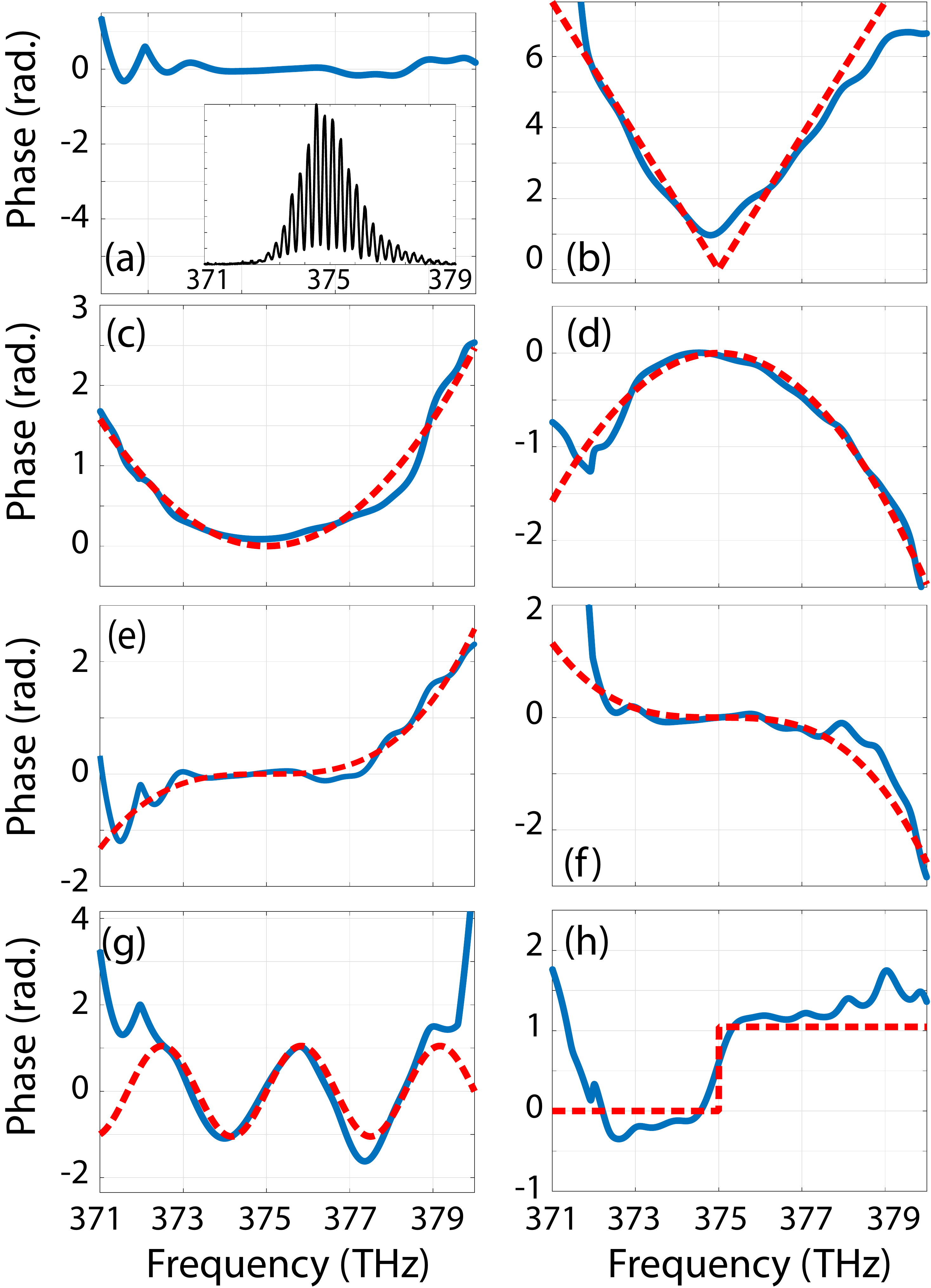}
 \caption{Characterization of various spectral phase $\varphi(\omega)$ applied with the pulse shaper: (a) unshaped with the interferogram in inset, (b) triangular phase function of $\pm300$ fs slope, (c-d) $\varphi(\omega)=\pm\phi^{(2)}/2(\omega-\omega_0)^2$ with $\phi^{(2)}=5~10^{3}$~fs$^2$, (e-f) $\varphi(\omega)=\pm\phi^{(3)}/6(\omega-\omega_0)^3$ with $\phi^{(3)}=5~10^{5}$~fs$^3$, (g) $\varphi(\omega)=\pi/3$ sin$(\omega\tau)$ with $\tau=300$ fs, (h) $\pi/3$ phase jump. The red dashed lines depicts the phase encoded with the pulse shaper and the blue lines the phase retrieved by our set-up.}
\label{Fig_charact_phases} 
\end{center}
\end{figure}

The first point to be checked is the ability of our device to generate suitable frequency shears $\Omega$ which constitutes the indispensable prerequisite for our purpose of pulse characterization. 
\begin{figure}[htpb]
\begin{center}
\includegraphics[width=\linewidth]{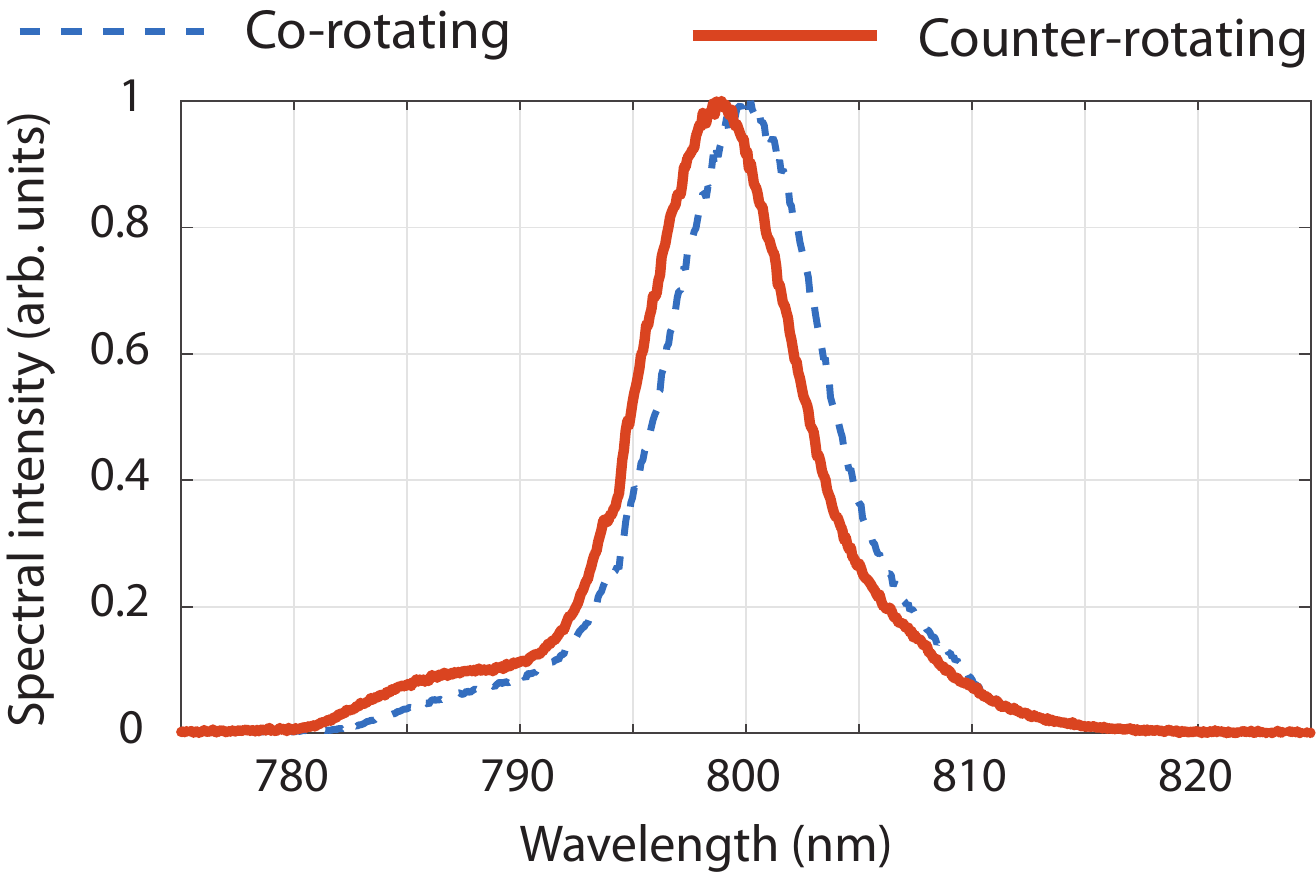} 
 \caption{Spectral shear produced by rotational Doppler shift.}
\label{FigShear} 
\end{center}
\end{figure} 
By rotating the last polarizer in front of the spectrometer (Figs.~\ref{FigDetection}(a,b)), one can choose to measure either the counter- or the co-rotating CP field. The measurement, shown in Fig.~\ref{FigShear}, reveals that the counter-rotating component is frequency shifted with respect to the co-rotating one as expected. The observed frequency shear $\delta\lambda\approx 1.1$~nm is in excellent agreement with the features of the TLP pulse produced by our polarization shaping. This undoubtedly points out the ability of the rotational Doppler effect to generate suitable shears for the characterization of fs laser pulse. When the polarizer (P) is turned so as to maximize the contrast of spectral fringes, the interferogram between the two components can be observed as shown in inset of Fig.~\ref{Fig_charact_phases}(a). The information about the spectral phase is in principle imprinted into the fringes pattern through the deviations from an equal frequency spacing. In order to verify this claim, several standard phase modulations have been applied using a pulse shaper device and then evaluated by our set-up via the inversion procedure given in Eq.~\ref{Eq_Phase_retrival} but evaluated at the mid-point for a better accuracy. The main results, summarized in Fig.~\ref{Fig_charact_phases}, demonstrate the ability of our device to retrieve different spectral phases with a great fidelity. As shown, standard cubic or quadratic spectral phases have been characterized (with their signs) together with more exotic modulations such as triangular or sinusoidal phase functions. 

We emphasize that the chirp to characterize should remain small enough to keep the main features of the TLP pulse unchanged (in order to preserve the Doppler shift). This condition, inherent to any SPIDER designs, is perfectly fulfilled for the quadratic and cubic phases investigated here. Our simulations \cite{paper_DERR2} reveal that for these phases, the TLP pulses is unaltered so that the Doppler effect provides an exact sheared replica of the input spectrum. For the sinusoidal and triangular phase functions, the alteration of the TLP pulse induces distortions of the spectral amplitude of the sheared pulse while its spectral phase is still well transferred. Since the relevant information is encoded along the fringes spacing and not along the amplitude, the phase retrieval matches the one applied as shown in Figs.~\ref{Fig_charact_phases}(b,g). Finally, simulations for the phase step [Fig.~\ref{Fig_charact_phases}(h)] predict that both spectral phase and amplitude of the field produced by Doppler effect are altered. Nevertheless, the deterioration of the transcribed phase remains moderate so that the phase retrieval remains satisfactory considering that commercial devices are generally designed to measure continuous and smooth spectral phases whereas phase jump are usually not covered. While this limitation is inherent to any SPIDER design (and to the most of self-referenced characterization methods), the ability of the present device to characterize this set of significantly different phases confirms the relevance of our DEER method. Furthermore, the use of alternative  strategies for producing the TLP pulse could provide a Doppler effect even more immune with respect to the input phase. The TLP pulse could for instance be produced by adding two counter-rotating CP fields with slightly different frequencies through the selection of two narrow sidebands within the spectrum. This could enable an improved characterization of phases with abrupt variations or jumps but at the expense of the energy consumed in the shaping process.
\section{Conclusion}
A new method of pulse characterization by spectral shearing interferometry has been proposed and implemented. The major innovation relies on the use of the rotational Doppler effect enabling the production of a frequency shear in the absence of frequency up-conversion. This non-conventional approach provides an extended functionality toward UV spectral domain and requires only one spectrometer. The performance of the set-up in terms of phase reconstruction is of high quality and confirms the relevance of the method. Several improvements of the present design can be envisioned for producing the sought-after pulse of rotating polarization. A thick dispersive medium could for instance reliably substitutes the pulse stretcher for characterizing broadband pulses (typically pulses whose FTL duration is below 30 fs). The use of pulse shaper devices could also be of great benefit to implement the polarization shaping with a better flexibility and compactness as compared to the present design. One challenge, in the topic of pulse characterization, is to find alternative nonlinear interactions for encoding in a robust way the spectral phase of short laser pulses. We believe that the rotational Doppler effect appears as a promising approach in this pursued objective as well as for other applications. This effect can for instance be applied to tune the wavelength of laser fields. In this context, the use of broadband radiations (such as super-continuum) to produce the TLP pulse should allow an ultrafast rotation of the field vector inducing controllable giant frequency shift. Doppler effect could be also of interest to control the wavelength of THz field  not otherwise easily accessible.

\section*{Funding}
This work was supported by the CNRS, the ERDF Operational Programme - Burgundy, the EIPHI Graduate School (Contract No. ANR-17-EURE-0002). This work was supported by a CIFRE PhD (2018/0253) granted by ANRT. Calculations were performed using HPC resources from DNUM-CCUB (Universit\'e de Bourgogne).\\
\bibliographystyle{apsrev4-2}

%

\end{document}